\documentclass[conference]{IEEEtran}
\IEEEoverridecommandlockouts
\newtheorem{theo}{Theorem}
\newtheorem{lemma}{Lemma}

\newtheorem{definition}{Definition}

\usepackage{cite}
\usepackage{amsmath,amssymb,amsfonts}
\usepackage{algorithmicx,algorithm}
\usepackage{algpseudocode}
\usepackage{graphicx}
\usepackage{textcomp}
\usepackage{xcolor}

\definecolor{maroon}{rgb}{0.5, 0, 0}
\definecolor{forestgreen}{rgb}{0.28, 0.49, 0.17}

\def\BibTeX{{\rm B\kern-.05em{\sc i\kern-.025em b}\kern-.08em
    T\kern-.1667em\lower.7ex\hbox{E}\kern-.125emX}}
    
\begin{document}

\title{An Improved Speedup Factor for Sporadic Tasks with Constrained Deadlines under Dynamic Priority Scheduling\\
{
}
\thanks{Work supported by NSFC (11101065), NSF (CNS 1755965).}
}

\author{
\IEEEauthorblockN{Xin Han, Liang Zhao}
	\IEEEauthorblockA{
		School of Software Technology\\
		Dalian University of Technology\\
		Email: hanxin@dlut.edu.cn, lightaini@qq.com
	}
\and
	\IEEEauthorblockN{Zhishan Guo}
	\IEEEauthorblockA{
		Department of Computer Science\\
		Missouri University of S\&T\\
		Email: guozh@mst.edu
	}
    \and
\IEEEauthorblockN{Xingwu Liu$^*$\thanks{$^*$Corresponding author}}
\IEEEauthorblockA{Institute of Computing Technology, CAS\\
University of Chinese Academy of Sciences\\
Email: liuxingwu@ict.a.cn}
}

 \maketitle
\thispagestyle{plain} \pagestyle{plain}
\begin{abstract}
Schedulability is a fundamental problem in real-time scheduling, but it has to be approximated due to the intrinsic computational hardness. As the most popular algorithm for deciding schedulability on multiprocess platforms, the speedup factor of partitioned-EDF is challenging to analyze and is far from been determined.
Partitioned-EDF was first proposed in 2005 by Barush and Fisher \cite{baruah2005partitioned}, and was shown to have a speedup factor 
at most $3-1/m$, meaning that if the input of sporadic tasks is feasible on $m$ processors with speed one, partitioned-EDF will always return succeeded on $m$ processors with speed $3-1/m$.
In 2011, this upper bound was improved to $2.6322-1/m$  by 
 Chen and Chakraborty \cite{chen2011resource}, and no more improvements have appeared ever since then.
In this paper, we develop a novel method to discretize and regularize sporadic tasks, which enables us to improve, in the case of constrained deadlines, the speedup factor of partitioned-EDF to $2.5556-1/m$, 
very close to the asymptotic lower bound $2.5$ in \cite{chen2011resource}. 
\end{abstract}




 \section{Introduction}

Scheduling is a hot topic in the  real-time systems community. Basically, given a finite set of tasks, each sequentially releasing infinitely many jobs, the mission of real-time scheduling is to allocate computing resources so that all the jobs are done in a timely manner. The fundamental question of schedulability naturally arises: Is it possible at all to successfully schedule these tasks, such that all of them receives enough execution before their deadlines? 

Unfortunately, answering this question is often not `easy'; e.g., the schedulability of a set for constrained-deadline sporadic tasks, which is the focus of this paper, is co-NP-hard even on a uniprocessor platform \cite{eisenbrand2010edf}.
For multiprocessor case, it remains NP-hard 
for partitioned paradigm, even if the relative deadlines of the tasks are required to equal their periods \cite{mok1983fundamental}. Here \emph{partitioned paradigm} means that once a task is assigned on a processor,
all the legal jobs released by the task
will be scheduled on the dedicated processor. These hardness results imply that it is almost impossible to exactly decide schedulability in polynomial time.

Due to the hardnesses, real-time schedulability problems are usually solved approximately by pessimistic algorithms which always answer `No' unless some sufficient-only conditions for schedulability are met.
To evaluate the performance of such an approximate algorithm (say, $\mathcal{A}$), the concept of {\em speedup factor}, also known as resource augmentation bound, has been proposed. Specifically, whenever a set of tasks is schedulable on a platform with speed one,  algorithm $\cal A$ will return `Yes' on the same platform with speed $r \ge 1$. The minimum such $r$ is referred to as the speedup factor of $\mathcal{A}$. Despite of some recent discussion on potential pitfalls \cite{chenLIPIcs2017} \cite{Guo_Dagstuhl2017} \cite{Awgural2018}, speedup factor has been a major metric and standard theoretical tool for assessing scheduling algorithms since the seminal work in 2000 \cite{KalyanasundaramP00}.

Recent years has witnessed impressive progress on finding schedulability decision algorithms with low speedup factors. For the preemptive case (i.e. running jobs might be interrupted by emergent ones ), Global-EDF has a speedup factor $2-1/m$ \cite{PhillipsSTW02} for scheduling set of tasks on $m$ identical processors, and there is a polynomial-time algorithm for uniprocessors
whose speedup factor is $1 + \epsilon$\cite{albers2004event}, where $\epsilon>0$ is sufficiently small
for uniform processors, refer to \cite{BaruahG08}.
For the non-preemptive case, there are also a variety of results, refer to  \cite{DevillersG00,DavisTGDPC18}.
Except the speedup factor,
there are many papers concerning about the utilization upper bound,
refer to \cite{BiniB05,Bini15,TheisF11}.

Although the speedup factor on uniprocessors is tight, the multiprocessor case remains open. Due to its simplicity, partitioned scheduling is of particular interest and has been attracting more and more attention from researchers and practitioners. Partitioned-EDF is the most popular schedulability decision algorithm of partitioned style, while the above-mentioned Global-EDF is not of partitioned paradigm. Breakthrough was made in the year of 2005, when Baruah and Fisher \cite{baruah2005partitioned} established a $4-2/m$ ($3-1/m, respectively$) upper bound
for the speedup factor of partitioned-EDF on arbitrary-deadline (constrained-deadline, respectively) task sets, where $m$ is the number of identical processors. A set of tasks is said to be constrained-deadline, if the relative deadline of each task is at most its period, otherwise is arbitrary-deadline. Then in 2011, Chen and Chakraborty \cite{chen2011resource} further improved  the speedup factor to $2.6322-1/m$ ($3-1/m$, respectively) for the constrained-deadline case (arbitrarily deadline case, respectively). Also in the same paper, a lower bound 2.5 of the speedup factor was established for the constrained-deadline case. Throughout the last seven years, the bounds in \cite{chen2011resource} were never improved.

It is worth noting that deriving the upper bound of the speedup factor of partitioned-EDF relies heavily on a quantity about scheduling on uniprocessors, denoted by $\rho$ which is formally defined in (\ref{eqn:rho}) of Section \ref{sec:pre}. Roughly speaking, $\rho$ measures how far the approximate demand bound function (defined in Section \ref{sec:pre}) deviates from the actual deadline time-points. Baruah and Fisher \cite{baruah2005partitioned}  bridged $\rho$ and the speedup factor of partitioned-EDF by showing that in case of constrained deadlines, the speedup factor is at most $1+\rho -1/m$. As a result, upper-bounding the speedup factor is reduced to upper-bounding $\rho$, and it is in this manner that both \cite{baruah2005partitioned} and \cite{chen2011resource} obtained their estimations of the speedup factor. Hence, the quantity $\rho$ itself deserves a deep investigation. Actually, Baruah and Fisher \cite{baruah2005partitioned} upper-bounded it by 2, Chen and Chakraborty \cite{chen2011resource}  narrowed its range into $[1.5,1.6322]$.

On this ground, this paper will explore a better upper bound of $\rho$, and on this basis, provide a better estimation of the speedup factor of partitioned-EDF for sets of constrained-deadline sporadic tasks. The contributions are summarized into the following three aspects.
\begin{enumerate}
\item  We improve the best existing upper bound of $\rho$ for constrained-deadline tasks from $1.6322$ to $1.5556$ (Theorem \ref{th:uniprocessor}), which is much close to the lower bound $1.5$. The speedup factor of partitioned-EDF  for the constrained-deadline case accordingly decreases from $2.632-1/m$  to $2.5556-1/m$ (Theorem \ref{th:multipro}), which is almost tight since there is an asymptotic lower bound $2.5$.
\item We find a way to losslessly discretize and regularize the constrained-deadline tasks so that essentially, the execution times of the tasks are all 1 and the deadlines are $1,2, \cdots,n$ respectively, where $n$ is number of tasks to be scheduled (Lemmas \ref{lemma:rational}, \ref{lemma:ei23}, \ref{lemma:mapping}). The only parameter that varies is the period. The transformation is lossless in the sense that the quantity $\rho$ does not change though the parameters are extremely simplified.
\item We invent a method to further transform the tasks so that the period of each task ranges over integers between 1 to $2n$ (Lemma \ref{lemma:dipidndi}). This transformation might be lossful, but the loss is negligible since $\rho$  changes at most 0.0556. These technics may be further applied to real-time scheduling or other problems.
\end{enumerate}

The rest of the paper is organized as follows:
Section 2 presents the model and preliminaries;
Section 3 focuses on uniprocessor case and derives 
 a new upper bound ($14/9$) of $\rho$
 for feasible sporadic tasks;
Section 4 provides a new upper bound ($23/9 - 1/m$) of the speedup factor for partitioned-EDF.
Finally, Section 5 concludes the paper and mentions some potential future directions.

\section{System Model and Preliminaries} \label{sec:pre}

We consider a finite set $\tau$ of sporadic tasks. Each task $\tau_i$ can be represented by a triple $\tau_i=(e_i, d_i, p_i)$,
where $e_i$ is the worst-case execution time,
$d_i$ is its relative deadline, and
$p_i$ is the minimum inter-arrival separation length (also known as period), respectively. The task $\tau_i$ is said to be constrained-deadline if $d_i\le p_i$. 

This paper focuses on constrained-deadline tasks. Hereunder, every task is constrained-deadline by default, unless otherwise mentioned. 

Given a task $\tau_i$,
we can calculate its demand bound function $dbf(\tau_i,t)$ \cite{baruah1990preemptively}  and its approximate demand bound function $dbf^{*}(\tau_i,t)$ \cite{albers2004event} in the following manner:
\begin{displaymath}
  dbf(\tau_i, t) =  \left \{
  \begin{array}{ll}
   0    &  \textrm{ if } t < d_i  \\
    \big(\left \lfloor \frac{t-d_i}{p_i}  \right \rfloor +1\big) \cdot e_i,  &  \textrm{ otherwise}
\end{array} \right.
\end{displaymath}
and
\begin{displaymath}
  dbf^{*}(\tau_i, t) =  \left \{
  \begin{array}{ll}
   0    &  \textrm{ if } t < d_i  \\
    (\frac{t-d_i}{p_i}  +1 ) \cdot e_i,  &  \textrm{ otherwise.}
\end{array} \right.
\end{displaymath}

Similarly for any set $\tau$ of tasks, we define
\[
  dbf(\tau, t) = \sum_{\tau_i \in \tau} dbf(\tau_i,t), \quad dbf^{*}(\tau, t) = \sum_{\tau_i \in \tau} dbf^{*}(\tau_i,t).
\]


To analyze the speedup factor of partitioned-EDF on multiprocessor platforms,
the following quantity plays a critical role:
\begin{align}
\rho = \sup_{\tau} \frac{dbf^*(\tau,d)}{d}, \label{eqn:rho}
\end{align}
where $\tau$ ranges over sporadic task sets that are feasible on uniprocessor platforms,
and $d$ is the largest relative deadline in $\tau$. Here \emph{feasible} means that the set of tasks allows a successful scheduling.

We will see that actually, $\rho$ is the optimum value of the following math programming $MP_0$:
  \begin{align}
 \sup & &  \frac{dbf^*(\tau, d_n)}{d_n},      \qquad \qquad \textrm{ ({\bf $MP_0$})}\\
subject\; to & &
 dbf(\tau,t) \le t,  \quad \forall t >0  \label{eqn:dbf0}\\
             & &    d_i + p_i > d_n, \quad   1 \le i  \le n-1, \label{eqn:dpdn0}
   \\
              && d_1  \le  d_2 \le \cdots  \le d_n,  \label{eqn:ddd0}\\
               && n\in \mathbb{Z}^+, e_i, d_i, p_i \in \mathbb{R}^+,  \quad 1 \le i  \le n.\label{eqn:epd0}
\end{align}
where $\mathbb{Z}^+$ is the set of positive integers while $\mathbb{R}^+$ stands for the set of positive real numbers.
\begin{lemma} \label{lemma:normalization}
$\rho$ is the optimum value of $MP_0$.
\end{lemma}
\begin{figure}
	\centering
	\includegraphics[width=0.45\textwidth]{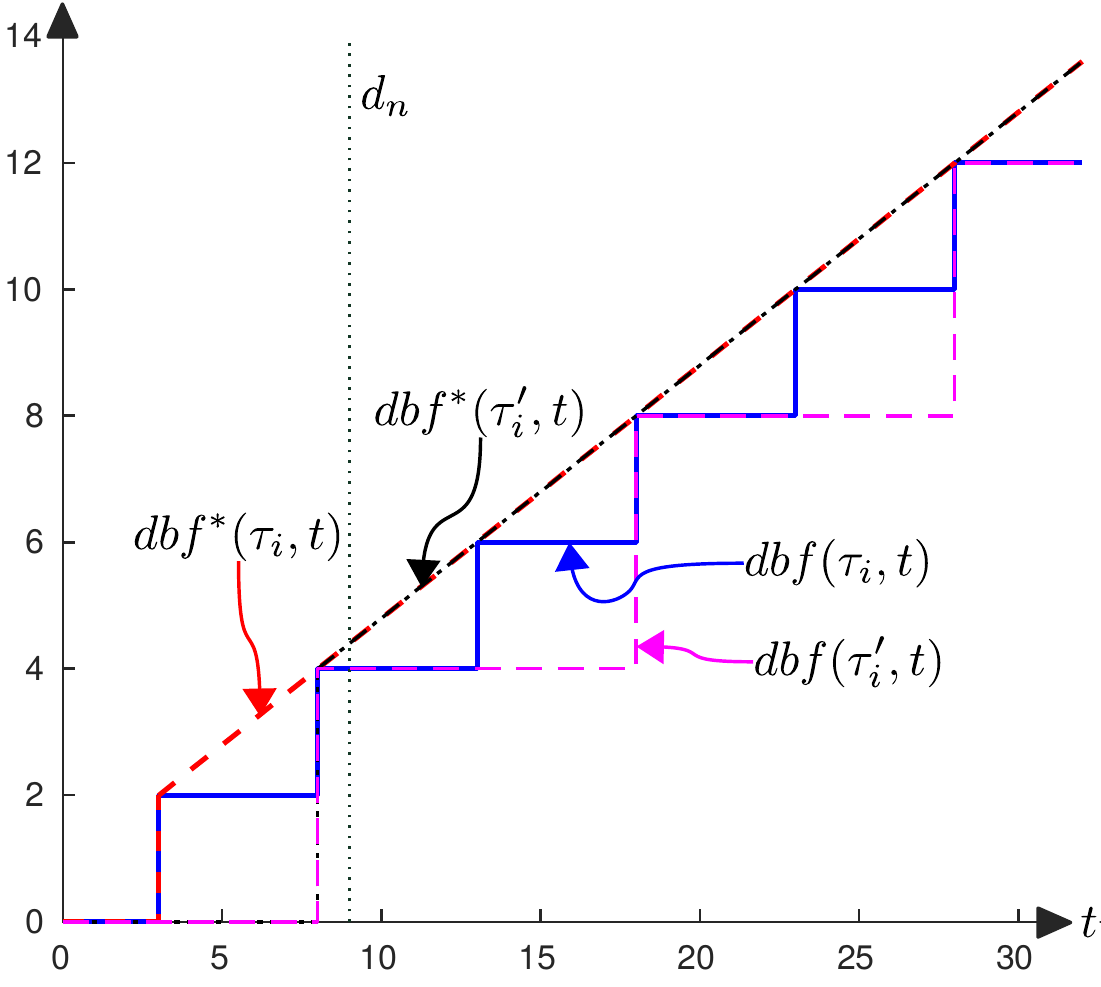}
	\caption{An example for task transformation and dbf modifications, with task parameters of $e_i=2, d_i=3, p_i=5,$ and $d_n = 9$.}
	\label{fig:utilitya}
\end{figure}
\begin{IEEEproof}
Let $\tau=\{\tau_i=(e_i,d_i,p_i):1\le i\le n\}$ be an arbitrary set of sporadic tasks that is feasible on a uniprocessor with speed 1. Assume that $d_1 \le d_2 \le \cdots \le d_n$. Apply the transformation proposed in \cite{chen2011resource}:

\begin{align}
    e_i^{'} &=& \left( \left\lfloor  \frac{d_n -d_i}{p_i}\right\rfloor +1\right) \cdot e_i, \\
    p_i^{'} &=& \left( \left\lfloor  \frac{d_n -d_i}{p_i}\right\rfloor +1\right) \cdot p_i, \\
    d_i^{'} &=& \left( \left\lfloor  \frac{d_n -d_i}{p_i}\right\rfloor \right) \cdot p_i + d_i.
\end{align}
Let $\tau^{'}=\{\tau_1^{'}, \tau_2^{'}, \cdots, \tau_{n}^{'}\}$ with $\tau_i^{'}=(e_i^{'},d_i^{'},p_i^{'})$ for any $1\le i\le n$.

Please refer to Figure \ref{fig:utilitya} for an illustration of the above mentioned transformation. 

For any $1\le i\le n$, $p_i^{'} -d_i^{'}=p_i - d_i$. Hence $\tau^{'}$ is constrained-deadline since so is $\tau$.

In \cite{chen2011resource}, it was proven that the following results hold simultaneously:

   i) $dbf^*(\tau, d_n)= dbf^*(\tau^{'}, d_n)$; 
   
   ii) $dbf(\tau, t) \ge  dbf(\tau^{'}, t)$ for $t >0$ ; 
   
  iii) $ d_n^{'} < d_i^{'} + p_i^{'}$ for $1 \le i \le n$; 
  
  iv)  $d_n ^{'} = d_n $.

This immediately leads to our lemma.
\end{IEEEproof}

\section{Improved Bound for Uniprocessor Case}
In order to estimate the speedup factor for multiprocessor partitioned scheduling, we first focus on the uniprocessor case.
The main result of this section is Theorem \ref{th:uniprocessor}, which establishes $14/9$ as an upper bound of $\rho$ for sporadic tasks.

The basic idea of our proof is to discretize the tasks into regular form, thus reducing the problem into an
optimization one on bounded integers. Roughly speaking, Lemma \ref{lemma:rational} makes sure that $\rho$ does not change if the parameters of the tasks are restricted to be rational numbers, Lemma \ref{lemma:ei23} claims that further requiring $e_i=d_i -d_{i-1}$ for all $i$ keeps $\rho$ unchanged, the trend continues by Lemma \ref{lemma:mapping} even if all the tasks are required to have the same worst-case execution time, and finally, Lemma \ref{lemma:dipidndi} enables us to only consider tasks with bounded discrete periods. These transformations reduce estimating $\rho$ to a simpler optimization problem which is solved approximately in Lemma \ref{lemma:14-9}. These results immediately lead to Theorem \ref{th:uniprocessor}.

Specifically, we first observe that the optimum value of $MP_0$ remains unchanged even if the domain $\mathbb{R}^+$ is replaced by $\mathbb{Q}^+$, the set of positive rational numbers.
 \begin{align}
 \sup & &  \frac{dbf^*(\tau, d_n)}{d_n},      \qquad \qquad \textrm{ ({\bf $MP_1$})}\\
subject\; to & &
 dbf(\tau,t) \le t,  \quad \forall t >0  \label{eqn:dbf1}\\
             & &    d_i + p_i > d_n, \quad   1 \le i  \le n-1, \label{eqn:dpdn1}
   \\
              && d_1  \le  d_2 \le \cdots  \le d_n,  \label{eqn:ddd1}\\
               &&  n\in \mathbb{Z}^+, e_i, d_i, p_i \in \mathbb{Q}^+,  \quad 1 \le i  \le n.\label{eqn:epd1}
\end{align}

  \begin{lemma} \label{lemma:rational}
  $MP_0$ and $MP_1$ has the same optimum value.
  \end{lemma}

  \begin{IEEEproof}
The lemma immediately holds if all of the following claims are true:
\begin{enumerate}
\item The objective functions of  $MP_0$ and $MP_1$ are the same and continuous.
\item The domain of $MP_1$ is included in that of $MP_0$.
\item For any $\epsilon>0$ and any feasible solution 
  $\tau=\{\tau_i=(e_i,d_i,p_i): 1 \le i  \le n\}$ to $MP_0$, 
     there is a feasible solution
     $\tau'=\{\tau'_i=(e'_i,d'_i,p'_i): 1 \le i  \le n\}$ to $MP_1$ such that for any $1 \le i  \le n$, 
     \begin{align}|e'_i-e_i|<\epsilon, |d'_i-d_i|<\epsilon, |p'_i-p_i|<\epsilon. \label{smallperturbation}\end{align} \label{perturbation}
\end{enumerate}

It suffices to prove Claim \ref{perturbation} since the others are obvious.

Let $\tau=\{\tau_i=(e_i,d_i,p_i): 1 \le i  \le n\}$ be an arbitrary set of tasks
that is a feasible solution to $MP_0$,
and $\epsilon$ be an arbitrary positive real number. Without loss of generality,
assume that $\epsilon<\min_{1 \le i  \le n} e_i$. For any $ 1 \le i \le n$,
arbitrarily choose
     \begin{align*}
p'_i\in (p_i+\frac{\epsilon}{2}, p_i+\epsilon)\cap \mathbb{Q}^+, \\
d'_i\in (d_i+\frac{(i-1)\epsilon}{2n}, d_i+\frac{i \epsilon}{2n})\cap \mathbb{Q}^+, \\
 e'_i\in (e_i-\epsilon,e_i)\cap \mathbb{Q}^+.
 \end{align*}
Let $\tau'$ denote the set of tasks $\{\tau'_i=(e'_i,d'_i,p'_i): 1 \le i  \le n\}$. Obviously, $\tau'$ meets Conditions (\ref{eqn:epd1}) and (\ref{smallperturbation}).

To proceed, arbitrarily fix an integer $1 \le i  \le n$.

Note that $p'_i-d'_i> p_i+\frac{\epsilon}{2}-(d_i+\frac{i \epsilon}{2n})\ge p_i-d_i$. This, together with the fact that $\tau_i$ is constrained-deadline, means $\tau'_i$ is also constrained-deadline.

Observe that
\[
d'_i>d_i+\frac{(i-1)\epsilon}{2n}\ge d_{i-1}+\frac{(i-1)\epsilon}{2n}>d'_{i-1}.
 \]
 Hence, $\tau'$ satisfies Condition (\ref{eqn:ddd1}) of $MP_1$.

Because
\begin{align*}
d'_i + p'_i &>d_i+ \frac{(i-1)\epsilon}{2n} + p_i+\frac{\epsilon}{2} \\
&\ge d_i+ p_i+\frac{\epsilon}{2}\\
&>d_n+\frac{\epsilon}{2} \textrm{\quad(since $\tau$ satisfies (\ref{eqn:dpdn0}))}\\
&>d'_n,
\end{align*}
the task set $\tau'$ satisfies Condition (\ref{eqn:dpdn1}).

As to Condition (\ref{eqn:dbf1}), arbitrarily fix $t>0$. 

When $t<d'_i$, $ dbf(\tau'_i,t)=0\le dbf(\tau_i,t). $ 

When $t\ge d'_i$,
 because $p'_i > p_i, d'_i > d_i, e'_i < e_i$, we have
 \begin{align*}
 dbf(\tau'_i,t)&=\left(\left\lfloor \frac{t-d'_i}{p'_i}  \right \rfloor +1\right) \cdot e'_i \\
 &\leq \left(\left\lfloor \frac{t-d_i}{p_i}  \right \rfloor +1\right) \cdot e_i=dbf(\tau_i,t).
  \end{align*}

As a result, we always have $dbf(\tau',t) \le dbf(\tau,t)$. Then $\tau'$ satisfies Condition (\ref{eqn:dbf1}) since $\tau$ satisfies (\ref{eqn:dbf0}).

Altogether, $\tau'$ is a feasible solution to $MP_1$. 
\end{IEEEproof}



Now we present a technical lemma that will be frequently used.
   \begin{lemma} \label{lemma:dipidn}
  Suppose $d, p, d', p'\in \mathbb{R}^+$ are such that $d + p= d' + p'$ and $d > d'$. For any real number $t$,
  \[
  \frac{t -d'}{p'} > \frac{t -d}{p}\]
   if and only if $t < d + p $.
  \end{lemma}

  \begin{IEEEproof}
   Let $ \delta = d - d'=p'- p$.
      Then
       \begin{align*}
       &&
    \frac{t-d'}{p'} > \frac{t-d}{p} \\
    \Leftrightarrow &&
    p \cdot(t-d') > p' \cdot(t-d)  \\
    \Leftrightarrow &&
     p \cdot(t-d+ \delta) > (p +\delta) \cdot(t-d) \\
   \Leftrightarrow &&
      p \cdot \delta >  \delta \cdot(t-d) \\
   \Leftrightarrow &&
     p  >  t-d.   
      \end{align*}

  \end{IEEEproof}
Hereunder, let $d_0=0$. Then it is time to show that the optimum value of $MP_1$ remains unchanged even if we further require $e_i  = d_i -d_{i-1}$ for all $i \ge 1$. We define a new math programming
  \begin{align}
\sup & &   \frac{dbf^*(\tau, d_n)}{d_n},      \qquad \qquad  (MP_2) \\
subject\; to & &
 dbf(\tau,t) \le t,  \quad \forall t >0  \label{eqn:dbf-2}\\
             & &    d_i + p_i > d_n, \quad   1 \le i  \le n-1, \label{eqn:dpdn2}
  \\
              && d_i = e_i + d_{i-1},   \quad 1 \le i  \le n, \label{eqn:di2}\\
               && n\in \mathbb{Z}^+, e_i, d_i, p_i \in \mathbb{Q}^+,  \quad 1 \le i  \le n. \label{eqn:epd2}
\end{align}

\begin{lemma} \label{lemma:ei23}
$MP_1$ and $MP_2$ have the same optimum value.

  \end{lemma}
  \begin{IEEEproof}
For any feasible solution $\tau=\{\tau_i=(e_i,d_i,p_i): 1 \le i  \le n\}$ to  $MP_1$, define 
$M(\tau)\triangleq |\{i: 1\le i\le n, d_i \ne e_i + d_{i-1}\}|$. Obviously, $\tau$ to $MP_1$ is a feasible solution to $ MP_2$ if and only if $M(\tau)=0$. 

Consider the following proposition: for any feasible solution $\tau$ to $MP_1$ with $M(\tau)>0$, there is a feasible solution $\tau'$ to $MP_1$ such that $M(\tau')<M(\tau)$ and the objective value of $\tau'$ is at least that of $\tau$. If it is true, one can easily prove the lemma by iteratively applying the proposition. Hence, the rest of the proof is devoted to showing this proposition.

Arbitrarily fix a feasible solution $\tau=\{\tau_i=(e_i,d_i,p_i): 1 \le i  \le n\}$ to $MP_1$. Suppose $M(\tau)>0$.
Assume   $k$ is  the smallest index such that
$e_k \ne d_k - d_{k-1}$, meaning that $ e_i =d_i - d_{i-1}$ for all $i < k$. Then we have
 \begin{align}
 \sum_{i=1}^{k-1} e_i= d_{k-1} \label{eqn:ee<d}.
 \end{align}

  Since $  d_i+ p_i> d_n\ge d_k\ge d_i$
    for any $i<k$, one has 
  \begin{align*}
      \sum_{i=1}^k e_i\le \sum_{i=1}^k dbf(\tau_i, d_k)\le dbf(\tau, d_k)  \le d_k,
  \end{align*} where the last inequality holds because $\tau$ satisfies Condition (\ref{eqn:dbf1}). This, together with (\ref{eqn:ee<d}), leads to $e_k \le d_k - d_{k-1}$. By the assumption that $e_k \ne d_k - d_{k-1}$, we get
 \begin{align}
 e_k < d_k - d_{k-1}.\label{eqn:e<dd}
 \end{align}

Construct $\tau'=\{\tau'_i=(e'_i,d'_i,p'_i): 1 \le i  \le n\}$ where 
\[
d'_i=d_i,p'_i=p_i,e'_i=e_i
 \]
 for any $i\ne k$, and 
 \[
 e'_k=e_k, d'_k=d_{k-1} +  e_k, p'_k=d_k + p_k -d_k'.
  \]
  By (\ref{eqn:ee<d}) and (\ref{eqn:e<dd}), $\sum_{i=1}^{k} e_i=d_{k-1}+e_k=d'_k<d_k$.

Obviously, $M(\tau')=M(\tau)-1<M(\tau)$, and $\tau'$ is constrained-deadline since so is $\tau$.

Now we prove that $\tau'$ is a feasible solution to $MP_1$. Since $\tau$ satisfies Conditions (\ref{eqn:dpdn1})-(\ref{eqn:epd1}), so does $\tau'$. To show that Condition (\ref{eqn:dbf1}) is satisfied by $\tau'$, we arbitrarily choose $t>0$ and proceed case by case.

\textbf{Case 1}: if $t<d'_k$. Then 
       \begin{align*}
 &dbf(\tau', t) =\sum_{1 \le i  \le n} dbf(\tau'_i, t) \\
     &=\sum_{1 \le i  <k} dbf(\tau'_i, t)   \quad\textrm{(because $t<d'_j$ for $j\ge k$)}\\
     &=\sum_{1 \le i  <k} dbf(\tau_i, t)  \quad\textrm{(because $\tau'_i=\tau_i$ for $i< k$)}\\
     &\le dbf(\tau, t)\\
     &\le t  \quad\textrm{(because $\tau$ satisfies Condition (\ref{eqn:dbf1}))}.
      \end{align*}

\textbf{Case 2}: if $d'_k\le t<d_k$.
       \begin{align*}
 &dbf(\tau', t) =\sum_{1 \le i  \le n} dbf(\tau'_i, t) \\
     &=\sum_{1 \le i  \le k}\left(\left\lfloor \frac{t-d'_i}{p'_i}  \right \rfloor +1\right) \cdot e'_i\\
     &=\sum_{{1 \le i  \le k}} e'_i \quad\textrm{(because $d'_i + p'_i>t$ for any $i$)}\\
     &=\sum_{{1 \le i  \le k}} e_i=d'_k\le t.
      \end{align*}

\textbf{Case 3}: if $d_k\le t < d'_k + p'_k$. Then 
       \begin{align*}
 dbf(\tau'_k, t)  &=\left(\left\lfloor \frac{t-d'_k}{p'_k}  \right \rfloor +1\right) \cdot e_k\\
     &=e_k \quad\textrm{(because $d'_k<d_k\le t < d'_k + p'_k$)}\\
     &=\left(\left\lfloor \frac{t-d_k}{p_k}  \right \rfloor +1\right) \cdot e_k,
      \end{align*} where the last equality is due to $d_k\le t < d'_k + p'_k=d_k + p_k$.
      
      For any $i\neq k$, $dbf(\tau'_i, t)=dbf(\tau_i, t)$ since $\tau'_i=\tau_i$. 
      
      As a result, $dbf(\tau', t)=dbf(\tau, t)\le t$ because $\tau$ satisfies Condition (\ref{eqn:dbf1}).

\textbf{Case 4}: if $t\ge d'_k + p'_k$. 
Because 
\[
d'_k< d_k \textrm{ and } 
p'_k+d_k'=d_k + p_k,\]
 by Lemma \ref{lemma:dipidn}, 
 we have 
 \[
 \frac{t-d'_k}{p'_k}\le \frac{t-d_k}{p_k}.\]
  Then
       \begin{align*}
 &dbf(\tau', t) =\sum_{1 \le i  \le n} dbf(\tau'_i, t) \\
     &=\sum_{i\ne k}dbf(\tau'_i, t) + \left(\left\lfloor \frac{t-d'_k}{p'_k}  \right \rfloor +1\right) \cdot e_k \\
     &\le \sum_{i\ne k}dbf(\tau'_i, t) + \left(\left\lfloor \frac{t-d_k}{p_k}  \right \rfloor +1\right) \cdot e_k \\
     &=\sum_{i\ne k}dbf(\tau_i, t)+ dbf(\tau_k, t) \quad\textrm{(since $\tau'_i=\tau_i$ for $i\neq k$)}\\
     &=dbf(\tau, t)\le t\quad\textrm{(since $\tau$ satisfies Condition (\ref{eqn:dbf1}))}.
      \end{align*}

Altogether, $\tau'$ satisfies Condition (\ref{eqn:dbf1}), so it is a feasible solution to $MP_1$.

Finally, we show that 
\[
\frac{dbf^*(\tau, d_n)}{d_n}\le \frac{dbf^*(\tau', d'_n)}{d'_n}.
 \]

When $k<n$, we have $d'_n=d_n$, so it suffices to show $dbf^*(\tau, d_n)\le dbf^*(\tau', d'_n)$. 

By definition of $\tau'$, for any $i\neq k$, $dbf^*(\tau_i, d_n)=dbf^*(\tau'_i, d'_n)$. Furthermore, note three facts: 
\begin{enumerate}
\item $p'_k+d_k'=d_k + p_k $;
\item $d'_k< d_k$;
\item $d_n<d_k + p_k $ due to Conditions (\ref{eqn:dpdn1}).
\end{enumerate}
By Lemma \ref{lemma:dipidn}, these facts mean 
 \[
 \frac{d_n-d_k}{p_k}  \le\frac{d'_n-d'_k}{p'_k},\] which implies $dbf^*(\tau_k, d_n)\le dbf^*(\tau'_k, d'_n)$. 

As a result, $dbf^*(\tau, d_n)\le dbf^*(\tau', d'_n)$.

When $k=n$, we have $d'_n<d_n$. For any $i< n$, 
       \begin{align*}
 \frac{dbf^*(\tau_i, d_n)}{d_n}=&\frac{e_i}{d_n}\left(1+\frac{d_n -d_i}{p_i}\right) \\
 =&\frac{e_i}{p_i}\left(1+\frac{p_i -d_i}{d_n}\right) \\
 <&\frac{e_i}{p_i}\left(1+\frac{p_i -d_i}{d'_n}\right) \\
 =& \frac{e_i}{d'_n}\left(1+\frac{d'_n -d_i}{p_i}\right) \\
 =& \frac{e'_i}{d'_n}\left(1+\frac{d'_n -d'_i}{p'_i}\right) \quad\textrm{(because $\tau'_i=\tau_i$)}\\
  =&\frac{dbf^*(\tau'_i, d'_n)}{d'_n}
      \end{align*}
where the inequality is due to $d'_n<d_n$ and $p_i -d_i\ge 0$ (since $\tau$ is constrained-deadline). In addition, 
       \begin{align*}
 \frac{dbf^*(\tau_n, d_n)}{d_n}=\frac{e_n}{d_n} <\frac{e'_n}{d'_n}=\frac{dbf^*(\tau'_n, d'_n)}{d'_n}.
      \end{align*}
 Therefore, we also get $
\frac{dbf^*(\tau, d_n)}{d_n}\le \frac{dbf^*(\tau', d'_n)}{d'_n}
 $, as desired.
   \end{IEEEproof}

We will impose further constraint on $MP_2$, without changing the optimum value. As presented in the math programming $MP_3$, the constraint is that all the $e_i$'s are equal.
%
%
%
  \begin{align}
\sup & &   \frac{dbf^*(\tau, d_n)}{d_n},      \qquad \qquad  (MP_3) \\
subject\; to & &
 dbf(\tau,t) \le t,  \quad \forall t >0  \label{eqn:dbf-3}\\
            & &    d_i + p_i > d_n, \quad   1 \le i  \le n-1, \label{eqn:dpdn3}   \\
              && d_i = e_i + d_{i-1},   \quad 1 \le i  \le n, \label{eqn:di3}\\
              &&  e_i =  d_n/n,   \quad 1 \le i  \le n,  \label{eqn:ed3}\\
               &&  n\in \mathbb{Z}^+, e_i, d_i, p_i \in \mathbb{Q}^+,  \quad 1 \le i  \le n. \label{eqn:epd3}
\end{align}


\begin{lemma} \label{lemma:mapping}
$MP_2$ and $MP_3$ have the same optimum value.
\end{lemma}
\begin{IEEEproof}
 Let $\tau =\{\tau_i=(e_i, d_i, p_i): 1\le i\le n\}$ be an arbitrary feasible solution to $MP_2$. 
 Due to Condition (\ref{eqn:epd2}), we can choose $\delta\in  \mathbb{Q}^+$ such that
  \[
 k(i)\triangleq \frac{e_i}{\delta}
  \]
  is an integer for any $1\le i\le n$. Let $n'=\sum_{i=1}^{n} k(i)$.

For any $1\le l\le n'$, define task $\tau'_l=\left(e'_l,d'_l,p'_l\right)$  as below, where $1\le i\le n$ and $1\le j\le k(i)$ are such that $l=m(i,j)\triangleq j+\sum_{1\le h<i}k(h)$:
%
   \begin{align*}
   e'_l &=\delta,\\
  d'_l &=  d_{i-1} + \frac{j}{k(i)}(d_i -d_{i-1}) = d_{i-1} + j \delta, \\
   p'_l &=  p_i + d_i - d'_l.
\end{align*}
Let $\tau'(i) = \{\tau'_{m(i,j)}: 1\le j\le k(i)\}$  for any $1\le i\le n$, and $\tau' = \cup_{i=1}^{n} \tau'(i)$. Let $d'_0=0$.
Next we will prove that $\tau'$ is a feasible solution to $MP_3$.

First of all, for any $1\le i\le n$ and $1\le j\le k(i)$, let $l=m(i,j)$. We have $d'_l\le d_i$ and $ p'_l =  p_i + d_i - d'_l\ge p_i $. Thus, $\tau'$ is constrained-deadline because so is $\tau$.

Since $\tau'$ satisfies Conditions (\ref{eqn:dpdn3})-(\ref{eqn:epd3}) by definition, 
now investigate Condition (\ref{eqn:dbf-3}). Arbitrarily fix $t>0$ and proceed case by case.

\textbf{Case 1}: $t < d'_{n'}$. Let integer $l\ge 0$ be such that $d_l'\le t<d_{l+1}'$. Then
       \begin{align*}
 &dbf(\tau', t) =\sum_{1 \le r \le n'} dbf(\tau'_r, t) \\
=&\sum_{1 \le r \le l} dbf(\tau'_r, t)  \quad\textrm{(because $t<d_{l+1}'$)}\\
     =&\sum_{1 \le r  \le l}\left(\left\lfloor \frac{t-d'_r}{p'_r}  \right \rfloor +1\right) \cdot e'_r\\
     =&\sum_{1 \le r  \le l} e'_r =d'_l\le t
    \end{align*}
where the fourth equality holds due to the inequality $p'_r>t-d'_r$ which in turn follows from three facts:
\begin{enumerate}
\item For any $1\le i\le n$ and $1\le j\le k(i)$, we have
  \[
      p'_{m(i,j)}=  p_i + d_i-d'_{m(i,j)} \textrm{ by definition};
  \]
\item For any $1\le i\le n$, $p_i + d_i>d_n$ since $\tau$ satisfies Condition (\ref{eqn:dpdn2});
\item $d_n=d'_{n'}>t$.
\end{enumerate}

\textbf{Case 2}: $t \ge d'_{n'}$. It suffices to prove that \[
dbf(\tau'(i),t) \le dbf(\tau_i,t)\] for any $1\le i\le n$. Arbitrarily fix $1\le i\le n$.

Suppose  $t < d_i+p_i$.
We observe that 
       \begin{align*}
 &dbf(\tau'(i), t) =\sum_{j=1}^{k(i)} dbf(\tau'_{m(i,j)}, t) \\
=&\sum_{j=1}^{k(i)} \left( \left\lfloor \frac{t-d'_{m(i,j)}}{p'_{m(i,j)}} \right\rfloor +1 \right)\delta\\
     =& k(i)\delta \quad\textrm{(because $t < d_i+p_i=d'_{m(i,j)}+p'_{m(i,j)}$)} \\
     =& e_i \quad\textrm{(By definition of $k(i)$}) \\
     =& dbf(\tau_i, t)  \quad\textrm{(because $d_i\le t < d_i+p_i$)} 
    \end{align*}

Then consider $t \ge d_i+p_i$. For any $1\le j\le k(i)$, since $d_i > d'_{m(i,j)}$ and $ d_i+p_i=d'_{m(i,j)}+p'_{m(i,j)}$, Lemma \ref{lemma:dipidn} implies
$$\frac{t-d'_{m(i,j)}}{p'_{m(i,j)}} \le  \frac{t-d_i}{p_i} ,$$
Which further leads to 
       \begin{align*}
 &dbf(\tau'(i), t) =\sum_{j=1}^{k(i)} \left( \left\lfloor \frac{t-d'_{m(i,j)}}{p'_{m(i,j)}} \right\rfloor +1 \right)\delta\\
     \le& \sum_{j=1}^{k(i)} \left( \left\lfloor \frac{t-d_i}{p_i} \right\rfloor +1 \right)\delta \\
     =& \left( \left\lfloor \frac{t-d_i}{p_i} \right\rfloor +1 \right)e_i  \\
     =& dbf(\tau_i, t) 
    \end{align*}

Altogether, Condition (\ref{eqn:dbf-3}) is satisfied in both cases, so $\tau'$ is a feasible solution to $MP_3$.

The rest of the proof is to show that 
\[
dbf^*(\tau', d'_{n'}) \ge dbf^*(\tau, d_n).\]
 Note that \[
 d'_{n'}=d_n<p_i+d_i =d'_{m(i,j)}+p'_{m(i,j)}\textrm{ and }  d'_{m(i,j)}\le d_i\]
  for any $1\le i\le n,1\le j\le k(i)$.  Lemma \ref{lemma:dipidn} implies that
$$\frac{d'_{n'}- d'_{m(i,j)}}{p'_{m(i,j)}}  \ge  \frac{d_n- d_i}{p_i}.$$
Then for any $1\le i\le n$, 
we have 
       \begin{align*}
 dbf^*(\tau'(i), d'_{n'})&=\sum_{j=1}^{k(i)} \left( \frac{d'_{n'}- d'_{m(i,j)}}{p'_{m(i,j)}}+1 \right)\delta\\
     &\ge \left(\frac{d_n- d_i}{p_i} + 1 \right)e_i\\
     &=dbf^*(\tau_i, d_n).
    \end{align*}

Therefore, $dbf^*(\tau', d'_{n'}) \ge dbf^*(\tau, d_n)$.
\end{IEEEproof}

It is still hard to find a good upper bound of the optimum value of $MP_3$, partly because Condition (\ref{eqn:dbf-3}) is too strong and Condition (\ref{eqn:dpdn3}) is too weak. 
It has to be modified accordingly.

On the one hand, we relax (\ref{eqn:dbf-3}) by replacing the function $dbf(\cdot,\cdot)$ with $f(\cdot,\cdot)$: 
for any task $\tau_i =(e_i, d_i, p_i)$ and time $t >0$,
\begin{displaymath}
f( \tau_i, t) = \left\{ \begin{array}{ll}
dbf(\tau_i, t) & \quad \textrm{if $t < d_i+ p_i$}\\
2e_i & \quad \textrm{ otherwise}
\end{array} \right.
\end{displaymath}
Note that $f( \tau_i, t)\le dbf( \tau_i, t)$ always holds. The first argument of $f$ can be naturally extended to any set $\tau$ of tasks:
\[
 f( \tau,t) = \sum_{ \tau_i \in \tau} f( \tau_i,t).
\]

On the other hand, instead of (\ref{eqn:dpdn3}), we require that the set of tasks $\tau$ should be aligned, as defined below:
\begin{definition} \label{defi:algn}
Given a task set $\tau =\{\tau_i=(e_i, d_i, p_i): 1\le i\le n\}$, a permutation $\pi$ over $\{1,2,\cdots,n\}$ is called an \emph{aligning permutation} of $\tau$ if 
 \[
 d_{\pi(i)} + p_{\pi(i)} = d_n + d_i\]
  for any $1\le i\le n$. $\tau$ is said to be \emph{aligned} if it has an aligning permutation.
\end{definition}


%
%
%

%

We will show that the optimum value of $MP_3$ does not decrease after the modification. Specifically, define a new math programming, where the tasks are \emph{not} required to be constrained-deadline:
  \begin{align}
\sup & &  \frac{dbf^*(\tau, d_n)}{d_n},     \qquad \qquad  (MP_4) \\
subject\; to & &
 f(\tau,t) \le t,  \quad \forall t >0
   \label{eqn:dbf-4}\\
   && \tau \textrm{ is aligned}, \label{eqn:align4}\\
   && d_i = e_i + d_{i-1},   \quad 1 \le i  \le n, \label{eqn:di4}\\
              &&  e_i =  d_n/n,   \quad 1 \le i  \le n,  \label{eqn:edn4}\\
               && n\in \mathbb{Z}^+, e_i, d_i, p_i \in \mathbb{Q}^+,  \quad 1 \le i  \le n. \label{eqn:epd4}
\end{align}

\begin{lemma} \label{lemma:dipidndi}
The optimum value of $MP_3$ is not more than that of $MP_4$.
\end{lemma}

\begin{IEEEproof}
Arbitrarily choose a feasible solution $\tau=\{\tau_i=(e_i, d_i, p_i): 1\le i\le n\}$ to $MP_3$. Let $\pi$ be a permutation over $\{1,2,\cdots,n\}$ such that
 \begin{align}
d_{\pi(1)} + p_{\pi(1)} \le d_{\pi(2)} + p_{\pi(2)} \le ...\le d_{\pi(n)} + p_{\pi(n)}.\label{eq:sorting}
 \end{align}

For any $1\le i\le n$, construct a task $\tau'_i=(e'_i, d'_i, p'_i)$ where
  \[
  e'_i=e_i, d'_i=d_i, p'_i=d_n +d_{\pi^{-1}(i)}- d'_i.\] 
  Let $\tau'=\{\tau'_i: 1\le i\le n\}$.

We will show that $\tau'$ is a feasible solution to $MP_4$. Since Conditions (\ref{eqn:align4})-(\ref{eqn:epd4}) are satisfied by definition, it suffices to investigate Condition (\ref{eqn:dbf-4}). Let's first derive an inequality as tool.

For any $1\le i\le n$, let $j=\pi^{-1}(i)$, and we have
 \begin{align*}
                   &d_i + p_i\\
                   \ge& dbf(\tau, d_i + p_i)    \quad \textrm{ (since $\tau$ satisfies Condition (\ref{eqn:dbf-3}))}\\
                   =& \sum_{1\le l\le j} dbf(\tau_{\pi(l)}, d_i + p_i) \\
                   &+ \sum_{j< l\le n} dbf(\tau_{\pi(l)}, d_i + p_i)    \\
                   \ge& \sum_{1\le l\le j}2e_{\pi(l)} + \sum_{j< l\le n} e_{\pi(l)}\\
                   =& \frac{2jd_n}{n}+\frac{(n-j)d_n}{n}\\
                   =& d_n+d_j \quad\textrm{  (due to Conditions (\ref{eqn:di3}) and (\ref{eqn:ed3}))}, 
 \end{align*} where the second inequality is because $$d_i + p_i=d_{\pi(j)} + p_{\pi(j)}\ge d_{\pi(l)} + p_{\pi(l)} \textrm{ for any $l\le j$}$$ and $d_i + p_i>d_n\ge d_{\pi(l)}$ for any $l$.

 Hence, we have 
  \begin{align}
                  d_i + p_i \ge d_n+d_{\pi^{-1}(i)}  =d'_i + p'_i  \label{dp-ineq}
 \end{align} by definition of $\tau'$.
 
  Now we continue to prove $\tau'$ satisfies Condition (\ref{eqn:dbf-4}). For an arbitrary $t>0$, this can be done case by case.

\textbf{Case 1}: $t<d_n+d_1$. 
Then for any $1\le i\le n$,
   \begin{align*}
                  &d_i + p_i \\
                  \ge&d'_i + p'_i= d_n+d_{\pi^{-1}(i)}  \quad \textrm{ by (\ref{dp-ineq})}\\
                  \ge& d_n+d_1 >t              
 \end{align*}
  This, together with the definition of $\tau'$, implies that
 \[
f(\tau',t)=dbf(\tau',t)=dbf(\tau,t).
 \]
 Because $\tau$ satisfies Condition (\ref{eqn:dbf-3}), we have $f(\tau',t)\le t$.

\textbf{Case 2}: $t\ge d_n+d_1$. Choose the biggest $1\le i\le n$ such that $d_n+d_i\le t$. Then  for any $j>i$, 
\[
p'_{\pi(j)}+d'_{\pi(j)}=d_n+d_j> t >d_n\ge d'_{\pi(j)} \textrm{  } .
\]

Thus
 \begin{align*}
                   &f(\tau', t) \\
                   =& \sum_{1\le j\le i} f(\tau'_{\pi(j)}, t) + \sum_{i< j\le n} f(\tau'_{\pi(j)}, t)    \\
                   =& \sum_{1\le j\le i}2e'_{\pi(j)} + \sum_{i< j\le n} e'_{\pi(j)}\\
                   =& d_n+d_i\le t
 \end{align*}

Altogether, Condition (\ref{eqn:dbf-4}) is also satisfied.

Furthermore, for any $1\le i\le n$, by (\ref{dp-ineq}) and $d_i=d_i'$, we have  $p'_i\le p_i$. This, together with $e'_i=e_i, d'_i=d_i$ for any $1\le i\le n$, implies $dbf^*(\tau', d'_n) \ge dbf^*(\tau, d_n)$. As a result, $$\frac{dbf^*(\tau, d_n)}{d_n}\le \frac{dbf^*(\tau', d'_n)}{d'_n}.$$

The lemma thus holds.
\end{IEEEproof}
We present a technical lemma before going on.
\begin{lemma} \label{lemma:cauchys}
For any $x_1,x_2,\cdots,x_n \in\mathbb{R}^+$ such that 
\[
\sum_{i=1}^n x_i = n^2,\]
we have 
\[
\sum_{i=1}^{n} \frac{i}{x_i} \ge \frac{4n}{9}.\]
\end{lemma}
\begin{IEEEproof}
By Cauchy's Inequality,
 \begin{align*}
  (\sum_{i=1}^{n} \frac{i}{x_i}) (\sum_{i=1}^{n} x_i)   \ge  ( \sum_{i=1}^{n} \sqrt{i})^2.
     \end{align*}

Note that
  \begin{align*}
  \sum_{i=1}^{n} \sqrt{i} &= n^{\frac{3}{2}} ( \sum_{i=1}^{n} \frac{\sqrt{i}}{\sqrt{n}} \cdot \frac{1}{n})   \\
   &\ge  n^{\frac{3}{2}}  \sum_{i=1}^n\int_{\frac{i-1}{n}}^{\frac{i}{n}} \sqrt{x} dx \\
   &=  n^{\frac{3}{2}}  \int_{0}^1 \sqrt{x} dx = \frac{2}{3} n^{\frac{3}{2}}.
  \end{align*}

Therefore,
 \begin{align*}
  (\sum_{i=1}^{n} \frac{i}{x_i})   \ge  \frac{\frac{4}{9}n^3}{n^2}= \frac{4n}{9}.
     \end{align*}
%
%
%
\end{IEEEproof}

\begin{lemma} \label{lemma:14-9}
The optimum value of $MP_4$  is at most $\frac{14}{9}$.
\end{lemma}
\begin{IEEEproof}
Arbitrarily choose a feasible solution $\tau=\{\tau_i=(e_i, d_i, p_i): 1\le i\le n\}$ to $MP_4$. Let $\delta=\frac{d_n}{n}$. By Conditions (\ref{eqn:di4}) and (\ref{eqn:edn4}), 
 \[
 e_i=\delta \textrm{ and } d_i=i\delta
 \] for any $1\le i\le n$.

Let $\pi$ be an aligning permutation of $\tau$. Then we have
 \begin{align*}
 \sum_{i=1}^{n}  p_{\pi(i)}= \sum_{i=1}^{n} (d_n+d_i-d_{\pi(i)}) = nd_n =  n^2 \delta,
 \end{align*}
which implies $\sum_{i=1}^{n}  \frac{p_{\pi(i)}}{\delta}=n^2$. By Lemma \ref{lemma:cauchys},
\begin{align*}
  \sum_{i=1}^n \frac{i \delta}{p_{\pi(i)}} \ge \frac{4n}{9} .
 \end{align*}

Hence,
\begin{align*}
  \sum_{j=1}^n \frac{d_j + p_j -d_n}{p_j} &=\sum_{i=1}^n \frac{d_{\pi(i)}+ p_{\pi(i)} -d_n}{p_{\pi(i)}} \\
  &=\sum_{i=1}^n \frac{d_i}{p_{\pi(i)}} \quad\textrm{ (since  $\tau$ is aligned)}\\
  &=\sum_{i=1}^n \frac{i \delta}{p_{\pi(i)}} \ge\frac{4n}{9}.
 \end{align*}

%
%
As a result, 
  \begin{align*}
 dbf^*(\tau, d_n) &= \sum_{i=1}^{n} dbf^*(\tau_i, d_n) \\
               &= \sum_{i=1}^{n} \left(2 - \frac{p_i + d_i -d_n}{p_i}\right) e_i\\
               &= \sum_{i=1}^{n} \left(2 - \frac{p_i + d_i -d_n}{p_i}\right) \delta \\
               & \le  2n \delta - \delta\sum_{i=1}^{n}  \frac{p_i + d_i -d_n}{p_i}\\
               &\le 2n \delta - \frac{4n}{9} \delta=\frac{14}{9}d_n
 \end{align*}
 The lemma holds.
\end{IEEEproof}

We are ready to present one of the main results of this paper, which claims that the resource augmentation bound on a single processor is at most 1.5556.
\begin{theo} \label{th:uniprocessor}
$ \frac{dbf^{*}(\tau, d)}{d} \le \frac{14}{9}$ for any set $\tau$ of constrainted-deadline tasks such that $dbf(\tau, t) \le  t$ for all $t > 0$, where $d$ is the maximum relative deadline of the tasks in $\tau$.
\end{theo}
\begin{IEEEproof}
It follows from Lemmas \ref{lemma:rational}, \ref{lemma:ei23}, \ref{lemma:mapping},  \ref{lemma:dipidndi}, and \ref{lemma:14-9}.
\end{IEEEproof}

\section{Partitioned Scheduling on Multiprocessors}
This section is devoted to partitioning sporadic tasks on multiprocessors, where the tasks are assumed to have Constrained Deadlines. We adopted the algorithm of Deadline-Monotonic Partitioning in \cite{baruah2005partitioned}. It is presented in Algorithm \ref{alg:partition} to make this paper self-contained, where $e_i$ and $d_i$ stand for worst-case execution time and relative deadline of task $\tau_i$, respectively.

Basically, Algorithm \ref{alg:partition}  assigns tasks sequentially in the order of non-decreasing relative deadlines. Suppose $\tau(k)$ is the set of
tasks at processor $k$ after the first $i-1$ tasks has been assigned. Then task $\tau_i$ is assigned to the first processor (say, processor No. $k$) that can safely serve the task, namely $e_i+dbf^*(\tau(k),d_i)\leq d_i$.

Remember that we have upper-bounded
\begin{align}
\rho = \sup_{\tau} \frac{dbf^*(\tau,d)}{d},
\end{align}
where $\tau$ ranges over constrained-deadline sporadic task sets that are feasible on uniprocessors,
and $d$ is the largest relative deadline in $\tau$.

The following lemma is from references \cite{baruah2005partitioned} and \cite{chen2011resource}, so the proof is omitted. 
\begin{lemma} \label{lemma:rho+1}
The speedup factor of Algorithm 1 is  $1 + \rho-1/m$, 
where $m$ is the number of processors.
\end{lemma}

It is time to present the other main result of this paper.
\begin{theo} \label{th:multipro}
The speedup factor for Algorithm \ref{alg:partition} is at most $2.5556-1/m$.
\end{theo}

\begin{IEEEproof}
The theorem immediately follows from Theorem \ref{th:uniprocessor} and Lemma \ref{lemma:rho+1}.
\end{IEEEproof}


\begin{algorithm}[h]
\caption{Deadline-Monotonic Partitioning}
\hspace*{0.02in} {\bf Input:}
sporadic tasks $\tau=\{\tau_1,\ldots,\tau_n\}$ to be partitioned on $m$ identical unit-capacity processors;

\Comment{The tasks are indexed non-decreasingly according to their relative deadlines. For any $1\le k\le m$, let $\tau(k)$ denote the set of tasks assigned to the $k$th processor.}
\begin{algorithmic}[1]
\State $\tau(k) \leftarrow \emptyset$, for any $1\le k\le m$;
\For{$i=1$ to $n$}
¡¡¡¡
¡¡¡¡        \If{there exists $k$ such that $e_i+dbf^*(\tau(k),d_i)\leq d_i$}
¡¡
%

\State Choose the smallest such $k$;
\State $\tau(k) \leftarrow \tau(k) \cup \{\tau_i\}$;
            \Else \State \Return FAIL
¡¡¡¡        \EndIf

\EndFor

\State \Return feasible assignment $\tau(1), \tau(2), \ldots ,\tau(m)$
\end{algorithmic}\label{alg:partition}
\end{algorithm}


\section{Conclusion and Future Work}
In this paper,
we improve the upper bound of the speedup factor of partitioned-EDF from $2.6322-1/m$ to $2.5556-1/m$ for constrained-deadline sporadic tasks on $m$ identical processors, narrowing the gap between the upper and the lower bounds from 0.1322 to 0.0556. This is an immediate corollary of our improvement of the upper bound of $\rho = \sup_{\tau} dbf^*(\tau,d)/d$ from 1.6322 to 1.5556. The new upper bounds are very close to the corresponding lower bounds. 

Technically, our improvements root at a novel discretization that transform the tasks into regular forms without decreasing $\rho$. The discretization essentially makes all the tasks have fixed execution times and deadlines. The only parameter that varies is the period, which is highly restricted so as to range over the set $\{1,2,\cdots,2n\}$, where $n$ is the number of tasks to be scheduled. By this transformation, estimating $\rho$ is reduced to a much simpler optimization problem. We believe that this knack may work in other problems or scenarios. 

However, we have not yet proved that our transformation is equivalent. This means that the discretization might strictly enlarge $\rho$. The good news is that the incurred loss, if not zero at all, is guaranteed to be no more than 0.0556.

As to future directions, we conjecture that Theorems \ref{th:uniprocessor} and \ref{th:multipro} remain true if the constrained-deadline condition is removed. We also conjecture that our method can derive a 1.5 upper bound for $\rho$, thus closing the gap between the upper and the lower bounds. If this is the case, the speedup factor of partitioned-EDF is also fully determined, at least in the case of constrained deadlines.  



\section*{Acknowledgment}

The authors would like to thank Prof. Sanjoy Baruah from Washington University at St. Louis for the fruitful discussions.

\bibliographystyle{IEEEtran}
\bibliography{references}

%

\end{document}